\documentclass[aps,prl,reprint,groupedaddress]{revtex4-1}
\usepackage{graphicx}
\usepackage{csquotes}

\begin{document}


\title{Simultaneous investigation of shear modulus and torsional resonance of solid ${}^{4}$He}

\author{Jaeho Shin}
\author{Jaewon Choi}
\author{Eunseong Kim}
\email[Corresponding author: ]{eunseong@kaist.edu}
\affiliation{Center for Supersolid and Quantum Matter Research, KAIST, Daejeon 305-701, Republic of Korea}
\author{Keiya Shirahama}
\affiliation{Department of Physics, Keio University, Yokohama 223-8522, Japan}

\date{\today}

\begin{abstract}
We investigate the origin of a resonant period drop of a torsional oscillator (TO) containing solid ${}^{4}$He by inspecting its relation to a change in elastic modulus. To understand this relationship directly, we measure both phenomena simultaneously using a TO with a pair of concentric piezoelectric transducers inserted in its annulus. Although the temperature, ${}^{3}$He concentration, and frequency dependence are essentially the same, a marked discrepancy in the drive amplitude dependence is observed. We find that this discrepancy originates from the anisotropic response of polycrystalline solid ${}^{4}$He connected with low-angle grain boundaries by studying the shear modulus parallel to and perpendicular to the driving direction.
\end{abstract}


\maketitle

The change in the shear modulus $\mu$ and the dissipation $Q^{-1}$ of solid ${}^{4}$He at low temperature \cite{DayBeamish2007, DaySyshchenkoBeamish2009, DaySyshchenkoBeamish2010, SyshchenkoDayBeamish2010, HaziotRojasFeffermanEtAl2013, HaziotFeffermanBeamishEtAl2013, HaziotFeffermanSourisEtAl2013, FeffermanSourisHaziotEtAl2014, SourisFeffermanMarisEtAl2014} have been understood thoroughly by the Granato-Lucke (GL) model \cite{GranatoLuecke1956, TeutonicoGranatoLuecke1964, LueckeGranatoTeutonico1968}. According to this model, in a dislocation network, dislocations glides under applied stress, which leads to an additional strain field. This strain decreases the $\mu$ of a solid from its intrinsic value. However, the slip motion can be effectively damped by binding of dislocation segments with impurities at low temperatures. The pinning of dislocations is regulated by the finite binding energy, $E_{b}$, between dislocations and impurities in a solid, such that the pinning can be promoted only at sufficiently low temperatures. These weakly bound impurities on the dislocations are detached as a result of increasing temperature and/or external stress, which can be described by the Debye relaxation process and characterized by a relaxation time $\tau$ and an activation energy $E_{b}$ \cite{SyshchenkoDayBeamish2010, FeffermanSourisHaziotEtAl2014, SourisFeffermanMarisEtAl2014}. Solid ${}^{4}$He is a golden testbed for the GL model because the only impurities in solid ${}^{4}$He are an extremely low concentration of ${}^{3}$He atoms. The properties of dislocation in solid ${}^{4}$He, such as the average network length, dislocation density, and length distribution have been extensively studied by Balibar \textit{et al}. \cite{HaziotRojasFeffermanEtAl2013, HaziotFeffermanBeamishEtAl2013, HaziotFeffermanSourisEtAl2013, FeffermanSourisHaziotEtAl2014, SourisFeffermanMarisEtAl2014}.

Another interesting observation of solid ${}^{4}$He is that the resonant period of a torsional oscillator (TO) containing solid ${}^{4}$He decreases below 0.2 K \cite{KimChan2004a, KimChan2004, RittnerReppy2006, RittnerReppy2007, RittnerReppy2008, AokiGravesKojima2007, ClarkWestChan2007, PenzevYasutaKubota2008, WestLinChengEtAl2009, HuntPrattGadagkarEtAl2009, ChoiKwonKimEtAl2010, ChoiTakahashiKonoEtAl2010, KimChoiChoiEtAl2011}. This was initially interpreted as a result of the reduction of the rotational inertia of solid ${}^{4}$He, and considered as the appearance of a ‘putative’ supersolid phase.  Nevertheless, both the $\mu$ and TO response exhibited fundamentally identical dependences on the temperature, driving amplitude, frequency, and amount of ${}^{3}$He impurities \cite{DayBeamish2007, KimChoiChoiEtAl2011, KimXiaWestEtAl2008}. To investigate the underlying relationships between them, Kim \textit{et al}. \cite{KimChoiChoiEtAl2011} measured the shear modulus change ($\Delta$$\mu$) and the resonant period drop ($\Delta$prd) simultaneously by inserting a pair of flat piezoelectric transducers (PZT) into the center of a TO. Even though a similar temperature dependence was observed, the drive amplitude responses were different. When a large AC voltage was applied to a driving transducer, the $\Delta$$\mu$ of solid ${}^{4}$He at the center channel was fully suppressed, but the shift in the resonance period of the TO was not significantly affected. Similarly, the influence of a TO drive on the $\mu$ measurements was also minor; the $\mu$ did not deviate from the unaffected value until the TO drive suppressed all of the non-classical responses of the TO. To explain this discrepancy in the drive amplitude dependence, the authors  suggested that there were ultimately different microscopic origins between the two phenomena. However, one could question the validity of this interpretation since the measurements were performed in different locations in the TO cell; hence, the discrepancy could be attributed to the different solid samples. 

Here, we present a new TO design in order to overcome the abovementioned problems by utilizing a pair of concentric PZT inserted into the annular channel of a TO cell, as shown in Fig. 1. This allows us to measure the TO response and change in $\mu$ originating from the same solid sample. The design of a TO capable of simultaneous measurements is not straightforward because it is necessary to selectively eliminate the influence of the complex geometry of a TO, and to directly associate the change in $\mu$ to the response of a TO. Recent studies \cite{KimChan2012, KimWestEngstromEtAl2012, BeamishFeffermanHaziotEtAl2012, Maris2012, ReppyMiJustinEtAl2012, KimChan2014, ChoiShinKim2015} have indicated that the inappropriate design of a TO would amplify the elastic effect of solid ${}^{4}$He. The analysis of the response in a non-ideal TO is not simple and is often misinterpreted. The discrepancies found in previous simultaneous measurements \cite{KimChoiChoiEtAl2011} can also be attributed to the complicated structure of the TO. Accordingly, it is crucial that the response of a TO should not be associated with the change in the $\mu$ in a complicated way, so that one can conclude clearly whether or not both phenomena are directly connected.
\begin{figure}
\includegraphics[width=0.37\textwidth]{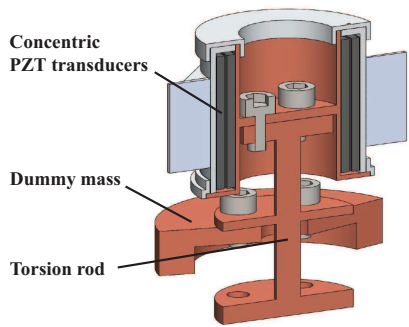}
\caption{Double-frequency torsional oscillator containing a pair of concentric piezoelectric transducers.}
\label{fig:01}
\end{figure}

We constructed a TO without any internal structure in order to eliminate a potential source of complexity in the analysis of the TO response, as shown in Fig. 1. The Maris effect \cite{Maris2012} in the current TO design is suppressed since the TO has the same topological structure as a torus constructed with a thick Be-Cu plate connected to a torsion rod by stainless steel screws. On the other hand, the “glue” effect \cite{ReppyMiJustinEtAl2012, KimChan2014, ChoiShinKim2015} is caused by relative motion between the components of a TO cell, which can be enhanced further when a TO is not rigidly constructed. Solid ${}^{4}$He in the narrow annular channel of the current TO is expected to exhibit the glue effect by consolidating the motion of the inner and outer TO walls at low temperatures. Since solid ${}^{4}$He is sandwiched between the inner and outer PZTs, it is possible to measure the change in $\mu$, which is directly coupled to the TO response. For small changes in the elastic modulus, a quantitative change in the resonance period can be simulated using a finite element method (FEM) simulation. FEM simulations enable us to obtain the optimum design of a TO in order to maximize the simple coupling of the change in $\mu$ to the TO response.

Besides, to confirm the origin of the TO response experimentally, we constructed a double-frequency TO so that the elastic effect can be examined via frequency analysis \cite{ChoiShinKim2015, MiReppy2013}. The resonance frequency of the lower mode ($f_{-}$) is 548 Hz and that of the higher mode ($f_{+}$) is 1280 Hz. The mechanical Q values are $1.99\times 10^{6}$ for the lower mode and $4.08\times 10^{5}$ for the higher mode. We grew solid ${}^{4}$He using  the conventional blocked capillary method, which is known to produce polycrystalline samples consisting of numerous randomly oriented micrometer-sized grains. We studied solid ${}^{4}$He grown with various ${}^{3}$He impurity concentrations of 0.6, 5, 10, 20, 75, 150, and 300 ppb. The ratios of the period reduction , $\Delta$prd, to the solid mass loading, $\Delta$P, $\Delta$prd/$\Delta$P, were approximately 0.5\% for the lower mode and approximately 3\% for the higher mode. The shift is larger than the values from the rigid TO experiments \cite{KimWestEngstromEtAl2012, KimChan2014, ChoiShinKim2015} and consistent with the values obtained from the FEM simulations.

\begin{figure}
\includegraphics[width=0.38\textwidth]{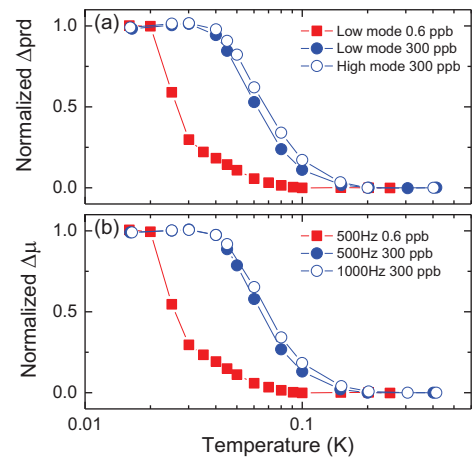}
\caption{(Color online) Temperature, ${}^{3}$He concentration and frequency dependence of (a) resonance period of TO and (b) shear modulus. Square (red) and circle (blue) symbols indicate the temperature behaviors for 0.6 and 300 ppb samples. Closed and open symbols represent the operating frequency.}
\label{fig:02}
\end{figure}

The temperature dependences of the lower and higher modes of the TO and the changes in $\mu$ are shown in Fig. 2. The changes in $\mu$ and the TO responses measured under various conditions are plotted together as a function of temperature for direct comparison. With isotopically pure ${}^{4}$He, the intermediate crossover temperature, $T_{i}$, from the low-temperature stiffened state to the high-temperature relaxed state was found to be approximately 27 mK in both measurements. We found $T_{i}$ increased with increasing ${}^{3}$He concentration and reached approximately 60 mK with commercially available ${}^{4}$He with a nominal ${}^{3}$He concentration of 300 ppb. This demonstrates that the temperature and ${}^{3}$He concentration dependences in both measurements are clearly identical. When the TO and PZT were driven at a higher frequency, $T_{i}$ was shifted to a higher temperature, as expected in the framework of dislocation pinning by ${}^{3}$He impurities.

\begin{figure}
\includegraphics[width=0.4\textwidth]{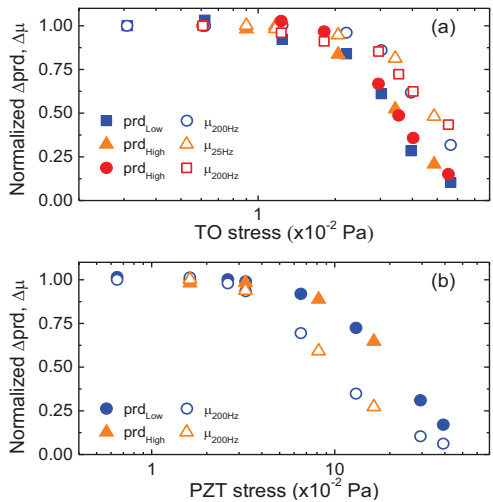}
\caption{Normalized $\Delta$prd (open symbols) and $\Delta$$\mu$ (closed symbols) are shown as a function of applied stress at 18 mK from temperature sweeps. (a) The influence of TO motion on resonance period and shear modulus measurements. (b) The influence of shear motion of PZT  on both measurements. Different colors indicate different frequencies.}
\label{fig:03}
\end{figure}

The most striking observation is that the drive dependence for both measurements shows an apparent discrepancy. In order to study the drive dependence, the $\mu$ and resonant period are each measured as a function of temperature with various frequencies and driving amplitudes. Then, $\Delta$$\mu$ and $\Delta$prd at 18 mK were extracted. These data at various measurement frequencies were normalized to highlight the drive dependence more clearly, and are plotted as a function of applied stress, as shown in Fig. 3. Because the elasticity of solid ${}^{4}$He can be altered by two independent driving sources, the PZT drive and the TO drive, two sets of data were individually collected to investigate the effect of drive dependence. Both the PZT and TO responses to the TO driving stress were monitored while maintained the PZT driving amplitude to a minimum in order to prevent its undesired additional influence. The apparent discrepancy is stood out when the drive dependences of the two phenomena are plotted together in Fig. 3(a). The TO stress value was calculated from the oscillation amplitude. Although both show a qualitatively the similar response, the quantitative dependence on the driving amplitude was not identical. No change in the $\mu$ measurements was observed until the magnitude of the driving stress increased to 0.02 Pa, where the TO response exhibited strong suppression by 15\% of the entire reduction of the resonant period. The TO responses for two separate frequencies follows the same drive dependent, whereas the $\mu$ traces a distinctly different path with the dependence shifted to the higher drive side.    
Similarly, the PZT-induced stress dependence is measured by holding the TO amplitude at a minimum. The stress caused by the PZT was converted from the driving voltage. The normalized drive dependences of both responses to PZT-induced stress are shown in Fig. 3(b). In contrast to the previous set of TO stress measurements, increasing the PZT drive causes a suppression of $\mu$ first without changing the TO response  while the current simultaneous measurements performed on the same solid sample. The threshold stress to induce suppression for the $\mu$ measurement is approximately 0.1 Pa, and that for the TO measurement is approximately 0.03 Pa.
\begin{figure}
\includegraphics[width=0.4\textwidth]{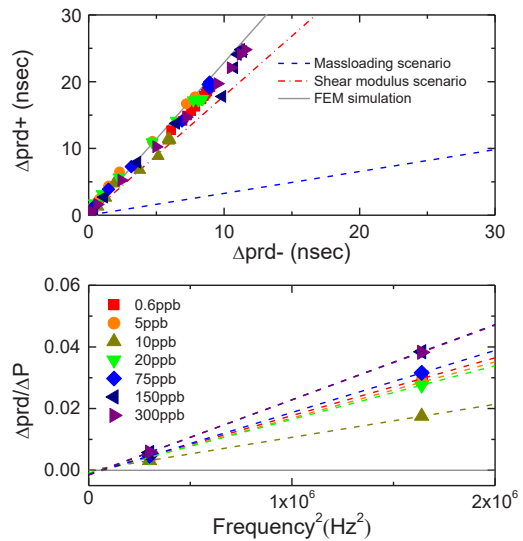}
\caption{(a) Ratio of $\Delta$prd for higher (+) to lower mode (-) with different ${}^{3}$He concentration. (b) Result of frequency analysis for frequency-independent contribution.}
\label{fig:04}
\end{figure}

Accordingly, it is very tempting to attribute the discrepancy in the simultaneous measurements to the different microscopic origins of the two phenomena, suggesting that the TO anomaly is possibly due to the ‘putative’ supersolid phase. However, a number of recent studies \cite{KimWestEngstromEtAl2012, KimChan2012, KimChan2014, ChoiShinKim2015} have strongly disproved the supersolid explanation. These TO studies demonstrate clearly that the TO anomaly does not exist in the ideal TO \cite{KimChan2012} or that the supersolid fraction should be less than 4 ppm\cite{KimChan2014}, indicating that the large TO anomaly was mainly due to the elastic effect of solid ${}^{4}$He. Recently, Reppy \textit{et al}. \cite{MiReppy2013} proposed that the frequency analysis of the TO could be used to distinguish the true supersolid response from the elastic effect of solid ${}^{4}$He. The period drop of an ideal TO measuring the change in rotational inertia due to a supersolid phase should be independent of the TO frequency. On the other hand, the TO response induced by the change in elastic modulus was expected to be linearly proportional to the square of the resonancet frequency, ${f}^{2}$. Reppy \textit{et al}. claimed to have measured the frequency-independent signal and interpreted it as the ‘true’ supersolid signal. However, Choi \textit{et al}. \cite{ChoiShinKim2015} investigated a frequency analysis using a rigid double torus-shaped oscillator and found no evidence of a supersolid. The upper limit of the supersolid signal from the frequency analysis of this study was 4 ppm, similar to the results of Kim \textit{et al}. \cite{KimChan2014}. 

Furthermore, we were able to determine experimentally whether or not the period reduction could be entirely explained by the change in elastic modulus of solid ${}^{4}$He. Figure. 4 (a) shows how the ratio, $\Delta$prd${}_{+}$/$\Delta$prd${}_{-}$, of the period drop of the higher mode ($\Delta$prd${}_{+}$) to the period drop of the lower mode ($\Delta$prd${}_{-}$) evolves under the various conditions. When the change in elastic modulus of solid ${}^{4}$He is the underlying mechanism for both anomalies, the ratio shows a steeper slope than expected in the ideal mass decoupling of solid ${}^{4}$He. This is because the change in period due to the elastic effect is proportional to the  ${f}^{2}$. Accordingly, the slope of the elastic effect can be steeper by a factor of $({f_{+}/f_{-}})^{2}$ than that of the supersolid scenario. Figure. 4 (a) shows that the slope for the elastic effect obtained from the frequency analysis (red dashed-dotted line), that of the FEM simulation of the elastic effect (gray line), and that expected from the supersolid explanation of mass decoupling (blue dashed line). The experimental data (closed circles) shows good agreement with the elastic effect obtained in the measurements and simulations. Furthermore, we could extract the frequency-independent superfluid contribution by subtracting ${f}^{2}$-dependent terms from the double-frequency TO results. Nearly zero mass decoupling is observed, as shown in Fig. 4(b), which is consistent with the implications of the ratio analysis. Based on both analyses, we concluded that the resonance period change of the TO anomaly is intrinsically originated from the elastic effect of solid ${}^{4}$He. 

We noticed that there were two key differences in the driving methods of the two measurements that might lead to the discrepancy in the drive dependence: the spatial profile of the strain along the driving direction and the orientation of the stress. First, the applied strain is uniform in the annulus when solid ${}^{4}$He is driven by a PZT \cite{DayBeamish2007}, while the strain due to torsional oscillation is expected to be parabolic with the maximum at the center of the annular channel and the minima at the confining annular walls \cite{Iwasa2013}. Despite the generic differences in the spatial profile, the same drive dependence in both measurements should have been observed. The time average of the change in $\mu$ is not very susceptible to the spatial space profile, but it is susceptible to the average stress applied to the solid ${}^{4}$He. Second, the directions of stress produced by both measurements were perpendicular to each other. The TO drive produced strain along the direction of torsional oscillation, whereas the concentric PZT induced strain and stress along the cylindrical axis of the TO cell, perpendicular to the TO drive. Nevertheless, these differences seem to be irrelevant to the explanation of the discrepancies.

The elastic modulus of a single crystal of solid ${}^{4}$He \cite{HaziotRojasFeffermanEtAl2013} can be quantitatively determined using the Bond matrix, which converts the elastic tensor, \textit{C}, in a crystal coordinate system to the elastic tensor, \textit{C'}, in a transducer coordinate system. The \textit{C} of a hexagonal close-packed crystal consists of five independent elastic components ($c_{11}$, $c_{12}$, $c_{13}$, $c_{33}$ and $c_{44}$) \cite{CrepeauHeybeyLeeEtAl1971, Greywall1977, DayBeamish2011}. A comprehensive study of an oriented single crystal of solid ${}^{4}$He instructed that the elastic anomaly of solid ${}^{4}$He is ascribed to the motion of dislocation on the Basal plane and can be described quantitatively by only the reduction in $c_{44}$ \cite{HaziotRojasFeffermanEtAl2013}. All 36 components in the elastic tensor \textit{C'} are generally non-zero and represent various values, depending on the relative angles between the crystal and transducer. Thus, the $\mu$ measurements of a single crystal of solid ${}^{4}$He reveal strong anisotropy. On the other hand, a polycrystalline solid ${}^{4}$He is composed of a sufficiently large number of grains with random orientations. The elastic tensor of a polycrystalline solid is isotropic and can be expressed by two independent values: the Young's modulus \textit{E} and the Poisson's ratio $\nu$. Thus, the discrepancies in the drive dependence would not be expected if the measurements were performed on a ‘proper’ polycrystal. Although the blocked capillary grown samples are considered to be a proper polycrystal, we attempted to confirm the validity of this hypothesis by measuring the stress that developed in both perpendicular and parallel to the driving direction.  

We constructed a new PZT-only cell in order to clarify the above issue. Two planar transducers on the detection side were stacked in such a way that the polarizing directions were perpendicular to each other. This enabled us to measure the $\mu$ on the parallel to the direction of the drive and that in the perpendicular direction \cite{Mulders2012}. On the driving side, two PZTs were stacked in the same way, so as to apply strain in both the perpendicular and parallel directions.
\begin{figure} [t]
\includegraphics[width=0.5\textwidth]{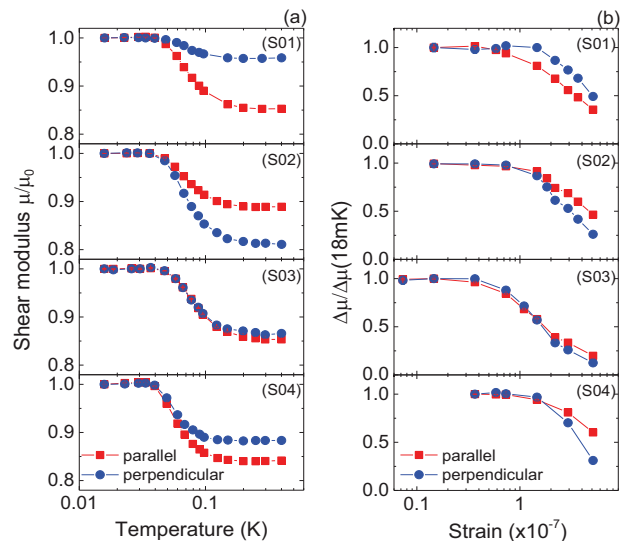}
\caption{(a) Temperature dependence and (b) driving amplitude dependence of shear modulus for different solid samples at 18 mK. Square symbols (red) indicate in a parallel direction measurement. Circle symbols (blue) indicate in a perpendicular direction measurement.} 
\label{fig:05}
\end{figure}
Figure. 5 shows the set of anisotropic responses for both PZTs during the measurements performed on four different samples that include two blocked capillary grown samples (S01, S02), an extensively annealed sample (S03), and a quench cooled sample (S04). Remarkably, the responses in both orientations clearly reveal the different drive dependences, despite the temperature dependences for each orientation being very similar. In addition, the anisotropic drive dependence remains essentially the same, regardless of the dramatically different sample preparation procedures. These anisotropic behaviors are not expected in a polycrystalline sample, as discussed earlier, revealing that the solid ${}^{4}$He between the transducers is composed of a few highly oriented crystals or a sufficiently small number of domains connected with a certain preferential orientation. While a small grain size of 10 $\mu$m or less was reported in a mass injected cell \cite{SasakiCaupinBalibar2008}, thermal conductivity \cite{ArmstrongHelmyGreenberg1979, ZmeevGolov2011} and X-ray diffraction measurements \cite{SchuchMills1962} reported the grain sizes of approximately 0.1 mm or larger. Assuming a grain size of 1 mm and, accordingly, approximately 20 grains between the transducers, the simulations qualitatively reproduced the prominent anisotropy. 

In summary, the unbinding of ${}^{3}$He impurities from dislocation lines is a fundamental mechanism to decrease the shear modulus of solid  ${}^{4}$He at high temperature and/or high stress. The temperature and  ${}^{3}$He dependence in both TO and shear modulus measurements can be understood straightforwardly in this framework. The reported anisotropy in the orientation-dependent drive response, on the other hand, has been a longstanding question and is often used as counter-evidence against a non-supersolid interpretation. Our sophisticated design of simultaneous TO and shear modulus measurements with the capability of frequency analysis enables us to understand the underlying connection deeply. The anisotropy arises from solid  ${}^{4}$He grown with a certain preferential orientation, indicating that both anomalies are originated from the same mechanism of a change in elastic property of solid ${}^{4}$He, rather than the emergence of supersolidity.
This work was supported by the National Research Foundation of Korea (NRF) grant funded by the Korean government (MSIP) (2007-0054-848).

\bibliography{Ref_Shin_2016}

\end{document}